\documentclass[ascmo, manuscript]{copernicus}
\usepackage{subfig}
\usepackage[ddmmyyyy]{datetime}
\usepackage{fancyhdr}
\usepackage{soul}
\usepackage[]{hyperref}
\usepackage[acronym]{glossaries}

\newacronym{3dvar}{3DVar}{three-dimensional variational}
\newacronym{bufr}{BUFR}{Binary Universal Form for data Representation}
\newacronym{carra}{CARRA}{Copernicus Arctic Regional Reanalysis}
\newacronym{cds}{CDS}{Copernicus Climate Data Store}
\newacronym{cerra}{CERRA}{Copernicus European Regional Reanalysis}
\newacronym{danra}{DANRA}{DANish regional atmospheric ReAnalysis}
\newacronym{dmc}{DMC}{deep moist convection}
\newacronym{dmi}{DMI}{Danish Meteorological Institute}
\newacronym{ecmwf}{ECMWF}{European Centre for Medium-Range Weather Forecasts}
\newacronym{era5}{ERA5}{ECMWF Reanalysis v5}
\newacronym{gts}{GTS}{Global Telecommunications System}
\newacronym{hpcf}{HPCF}{High-Performance Computing Facility}
\newacronym{icdr}{ICDR v1.1}{Interim Climate Data Record}
\newacronym{lam}{LAM}{Limited Area Modeling}
\newacronym{mars}{MARS}{Meteorological Archival and Retrieval System}
\newacronym{mslp}{MSLP}{Mean Sea Level Pressure}
\newacronym{nckf}{NCKF}{Nationalt Center for Klimaforskning}
\newacronym{nwp}{NWP}{Numerical Weather Prediction}
\newacronym{oi}{OI}{optimal interpolation}
\newacronym{s10m}{S10m}{10\,metre wind speed}
\newacronym{synop}{SYNOP}{Surface Synoptic Observations}
\newacronym{t2m}{T2m}{2\,metre temperature}
\newacronym{wmo}{WMO}{World Meteorological Organisation}

\linenumbers

\begin{document}

\title{DANRA: The Kilometer-Scale Danish Regional Atmospheric Reanalysis}

% \Author[affil]{given_name}{surname}
\Author[1]{Xiaohua}{Yang}
\Author[1]{Carlos}{Peralta}
\Author[1]{Bjarne}{Amstrup}
\Author[1]{Kasper}{Stener Hintz}
\Author[1]{Søren Borg}{Thorsen}
\Author[1]{Leif}{Denby}
\Author[1]{Simon}{Kamuk Christiansen}
\Author[1]{Hauke}{Schulz}
\Author[1]{Sebastian}{Pelt}
\Author[1]{Mathias}{Schreiner}

\affil[1]{Danish Meteorological Institute}

\correspondence{Xiaohua Yang (xiaohua@dmi.dk)}

\runningtitle{DANRA}

\runningauthor{Xiaohua Yang}

\newdateformat{slashdate}{\THEDAY/\THEMONTH/\THEYEAR}

\date{Draft as of \slashdate\today}

\received{}
\pubdiscuss{} %% only important for two-stage journals
\revised{}
\accepted{}
\published{}

\newcommand\sectionaim[2]{\textit{\textbf{Aim}: #1, \textbf{Responsible}: #2}}

\firstpage{1}
\maketitle

\begin{abstract}
The \gls{danra} is a novel high-resolution (2.5 km) reanalysis dataset covering Denmark and its surrounding regions over a 34-year period (1990-2023). Denmark’s complex coastline, with over 400 islands and an extensive 7,400 km coastline, means that most municipalities experience mixed land-sea variability. This complexity requires a regional climate reanalysis that can resolve fine-scale coastal and inland features, as well as their impact on climate variability. \gls{danra} is based on the HARMONIE-AROME \gls{nwp} model and assimilates a comprehensive set of observations, with a particular focus on Denmark. Compared to global reanalyses such as the \gls{era5}, \gls{danra} demonstrates superior performance in representing essential climate variables, including near-surface weather parameters during both extreme and ordinary conditions. We illustrate these improvements in the representation of several extreme weather cases over Denmark, such as the December 1999 hurricane-force storm, the July 2022 national temperature record, and the August 2007 cloudburst in South Jutland. \gls{danra} is made to support climate adaptation, impact modelling, and the training of next-generation data-driven atmospheric forecasting models. \gls{danra} is distributed as Zarr dataset freely accessible from an object store, maximizing its usability for climate adaptation, impact modelling, and data-driven research.
\end{abstract}

\introduction
\glsresetall
Global reanalysis datasets are extensively utilized in climate research as well as planning and development of both private and public projects, providing valuable information on historical weather and climate. Recently, a rapidly growing machine learning community has leveraged global reanalysis data to develop data-driven models, which have shown promising results \citep{Price2025,lam2023}. This interest is driving an increasing demand for high-quality, state-of-the-art reanalysis datasets in both private and public bodies. However, compared to regional reanalyses based on \gls{lam}, global reanalysis datasets typically lack sufficient horizontal resolution to capture weather extremes. This limitation often restricts their applicability in climate studies and modeling activities, particularly for local (municipal-level) applications, renewable energy planning \citep[]{gualtieri2022} and the representation of extreme weather conditions. To address these shortcomings, several regional reanalysis datasets have been developed, offering higher resolution and improved local representation \citep[]{gleeson2017,carra2020,Køltzow_etal_2022,ridal2024}. 

\gls{danra} is a kilometer-scale regional reanalysis dataset, specifically designed to overcome the limitations of global reanalyses in regional applications. Denmark’s complex coastal geography, characterized by numerous fjords and over 400 islands, yields an extensive coastline of approximately 7,400 km \citep[]{Moller1988}. Given the country’s small size, this lengthy coastline significantly influences local weather and climate variability. Most Danish municipalities feature mixed land-water coverage, making high-resolution climate reanalysis—capable of accounting for coastal complexity—particularly valuable for regional planning and climate adaptation efforts. 

At its core, \gls{danra} leverages the state-of-the-art operational \gls{nwp} model at \gls{dmi}, HARMONIE-AROME, as its foundational system \citep{yang2017}. The reanalysis operates on a $800 \times 600 \times 65$ grid, featuring a grid-spacing of $2.5$ km.
With a reanalysis system based on a non-hydrostatic, mesoscale convection-resolving weather forecasting system, \gls{danra} achieves a kilometer-scale spatial resolution, which is unprecedented among reanalysis products for the region. This poses \gls{danra} as a promising high-fidelity dataset for climate applications in Danish society.

Reanalysis datasets typically span several decades, reflecting the evolution of observation systems as well as climate conditions. Choosing a state-of-the-art operational forecasting system for reanalysis helps secure high-quality data while making production more stable, efficient, and maintainable. For the \gls{danra} reanalysis, the system base is adapted from the \gls{carra} \citep{carra2020} system. The \gls{carra} system is based on the operational \gls{nwp} HARMONIE-AROME Cycle 40h1 assimilation and forecast model systems used in the national weather services of Denmark and Norway \citep{bengtsson2017} for weather prediction over the European Arctic regions. Among the adaptations made for the \gls{carra} reanalysis, the focus is mainly on introducing and adapting additional observation data streams and improving modelling for the 'cold' climate condition in the Nordic and Arctic regions, particularly for ice and snow conditions. In \gls{danra}, the primary focus of the adaptation from \gls{carra} is the inclusion of additional observation data collected over Denmark and nearby regions, and harmonisation of configuration parameters to be consistent with those used in the operational \gls{nwp} system at \gls{dmi} for the Danish area \citep{yang2017}. For the lateral boundary, \gls{danra} utilizes the reanalysis data from the Copernicus Climate Change Service (C3S) global reanalysis \gls{era5} \citep{hersbach2020era5}. 

In this paper, we present a comprehensive 34-year dataset of the \gls{danra} reanalysis (1990-2023). The reanalysis encompasses a wide range of weather events, offering a unique opportunity to test the system's ability to replicate various meteorological conditions, including weather extremes. In assessing the quality of the \gls{danra} reanalysis dataset, we have benchmarked it against \gls{era5}, which is widely recognized as the reference for reanalysis in recent decades at the global scale. This comparative approach highlights the value of resolution and detail, which is a pivotal aspect in understanding and predicting weather phenomena crucial to the regional climate. The insights gained from this evaluation show promise to significantly enhance our comprehension of weather patterns and trends, particularly in the context of Denmark's unique climatic conditions.
During the pilot phase of the DANRA project, \gls{danra} reanalysis system has been examined for a series of high-impact weather episodes to demonstrate the prediction capability in representing Danish weather conditions, including extremes \citep{yang2021}. A case study is presented for each historical extreme in temperature, wind, and heavy precipitation in this paper. These studies underline the quality of the \gls{danra} high-resolution dataset. For comparison, the corresponding data from the Copernicus 5.5 km \gls{cerra} reanalysis, as well as the 31 km global reanalysis \gls{era5}, have also been extracted to examine the added value of higher-resolution reanalyses. 
The \gls{danra} dataset has been converted to CF-compliant Zarr data and made freely accessible in an S3-compatible object store. With this effort, we maximise the usability of the dataset for climate adaptation, impact modelling, data-driven research, and other stakeholders. Converting such a big dataset is non-trivial. Details about this are presented in section \ref{sec:zarr}.

\section{Methods}

\subsection{Reanalysis system}

In terms of model source, \gls{danra} employs \gls{dmi}'s operational HARMONIE-AROME-40h1.1 \gls{nwp} model suite \citep{yang2017}. The HARMONIE-AROME-40h1.1 model \citep{bengtsson2017} is a non-hydrostatic, convection-permitting \gls{nwp} system that has been used extensively by a wide array of European national weather services contributing to the ACCORD \gls{nwp} research collaboration (umr-cnrm.fr/accord/).

The \gls{danra} reanalysis operates through a cyclic process of analysis and short-range forecasting, updated every three hours at 00, 03, 06,... and 21 UTC. During the analysis step, the system assimilates both surface and upper-air observations using a \gls{3dvar} data assimilation scheme for upper-air model states, and multivariate \gls{oi} for the near-surface and soil properties. Each cycle is followed by a short-range forecast: 18 hours for the 00 and 12 UTC cycles, and 3 hours for all other analysis times. The extended forecast lead time at 00 and 12 UTC is designed to enable the extraction of accumulated precipitation through model integration beyond a 6-hour lead time. This approach mitigates potential moisture spin-up, a common issue during the initial stages of model integration following data assimilation. This configuration aligns with practices adopted in other reanalysis projects \citep{hersbach2020era5, ridal2024, gleeson2017}.

\begin{figure}[t]
\centering
   \includegraphics[width=14.5cm]{"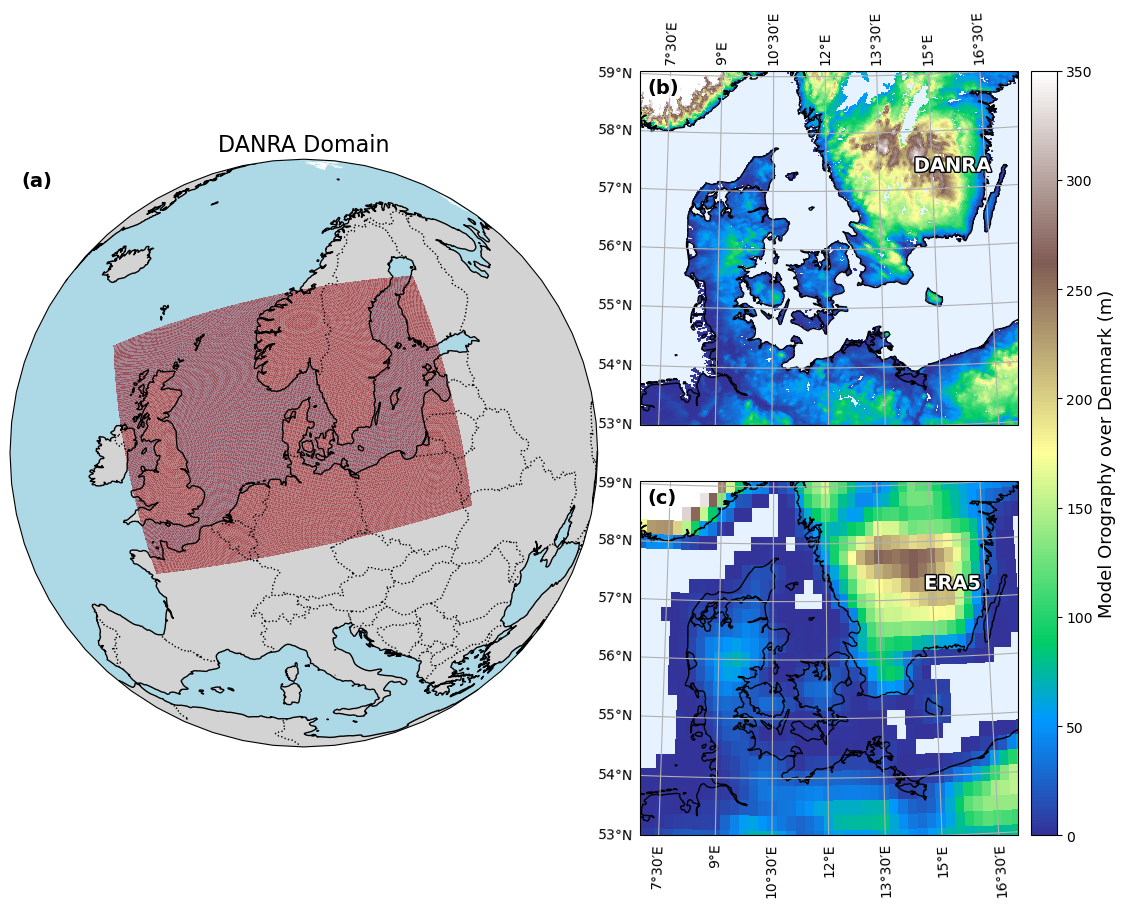"}
   \caption{(a) Domain of the \gls{danra} reanalysis, (b) Model orography over Denmark of \gls{danra} at 2.5 km horizontal resolution and (c) model orography over Denmark of \gls{era5} at 31 km horizontal resolution.}
    \label{fig:danra2014}%
\end{figure}

The \gls{danra} reanalysis is conducted at a horizontal $2.5$ km grid-spacing on the DKA model domain, the operational DMI-HARMONIE-AROME domain used for weather forecasting over Denmark during the period 2012–2016. The DKA domain consists of an 800 × 600 horizontal grid and 65 vertical levels, as illustrated in Figure \ref{fig:danra2014}a. To highlight the resolution advantage, the grid-averaged orography of \gls{danra}’s 2.5 km × 2.5 km grid is compared to that of \gls{era5}’s 31 km × 31 km grid \citep[][Figure \ref{fig:danra2014}b and \ref{fig:danra2014}c, respectively]{hersbach2020era5}. The comparison reveals a substantially improved representation of Denmark’s coastal landscape in the kilometer-scale \gls{danra} reanalysis compared to the coarser-resolution global \gls{era5} model. Higher resolution enhances quantitative realism in depicting local-scale weather and climate characteristics, making it particularly valuable for municipality-level climate adaptation planning. 

For reanalysis applications, the \gls{danra} system incorporates adaptations similar to those implemented in \gls{carra} \citep[][]{carra2020}, including the use of hourly \gls{era5} analyses at the lateral boundaries, as well as system modifications optimized for reanalysis. In the data assimilation process,  we include all conventional observations used by \gls{era5} at \gls{ecmwf} as a baseline, supplemented by additional local observations not archived by \gls{ecmwf} (see Section \ref{obs_and_qa}). The upper-air data assimilation further integrates an extensive array of satellite remote sensing data, such as microwave and infrared radiances, atmospheric motion vectors, scatterometer measurements, and radio occultation data. This represents a substantial expansion of observational input compared to the data usage in operational \gls{nwp} forecasts for the same historical periods, particularly during the early years of the \gls{danra} reanalysis.

The \gls{danra} reanalysis was produced using the \gls{hpcf} at \gls{ecmwf}, leveraging computational resources allocated to \gls{ecmwf} member states, with a total compute time of approximately 80,000 node hours.

\subsection{Observational data acquisition and quality assurance}\label{obs_and_qa}

Any reanalysis integrates observational data with \gls{nwp} models through data assimilation to produce a spatially and temporally continuous historical weather record. For surface assimilation in \gls{danra}, in-situ measurements, including screen-level temperature, humidity, and snow depth observations, are assimilated to update near-surface soil states. For upper-air assimilation, \gls{danra} incorporates surface pressure and upper-air wind, temperature, and humidity data collected from conventional in-situ platforms such as synoptic weather stations, ships, drifting buoys, aircrafts, and radiosondes, as well as satellite remote sensing data. For most conventional and remote sensing observations, the data used by the \gls{era5} reanalysis as archived by the \gls{mars} at \gls{ecmwf}, serve as the baseline. Additionally, \gls{danra} enhances observation coverage by assimilating reprocessed atmospheric motion vectors, satellite scatterometer wind data, and reprocessed GPS radio occultation bending-angle data from the ROM SAF \gls{icdr} \citep{ROMSAF_ICDR_v11}. 

As detailed in \citet{yang2021}, the \gls{danra} region has experienced significant changes in observational data availability over the past three decades. Aircraft measurement data became available in the late 1990s, while satellite remote sensing data increased steadily over the period from virtually non-existent. Additionally, in-situ surface observations have seen a marked expansion. For example, the availability of mean sea level pressure (MSLP) and screen-level temperature data more than doubled between the early 1990s and 2015, driven by both an increase in the number of stations and higher reporting frequencies.

A key feature of the \gls{danra} reanalysis is its comprehensive collection, rescue, and quality assurance of in-situ observation in and around Denmark. This effort builds on the extensive work of the Copernicus regional reanalysis projects \gls{carra} and \gls{cerra} (2018-2021), which aimed to maximize the availability of in-situ observations over the European Arctic and Pan-European reanalysis regions, respectively. One of the focuses in such efforts has been the collection of local observations that were previously not widely distributed via the \gls{gts}, the backbone of global meteorological observation exchange coordinated by the \gls{wmo}. 

For \gls{danra}, the enhancements on in-situ data encompass the collection of surface observations previously outside of the \gls{gts}, as well as stations reporting at higher frequencies than those available through the \gls{gts}. For example, while \gls{gts}-distributed observations in the 1990s - the start of the \gls{danra} reanalysis period — were typically limited to 3- to 6-hour intervals, \gls{danra} leverages hourly \gls{synop} records for most Danish stations throughout its coverage. In fact, beginning around 1998, a rapidly expanding network of Danish \gls{synop} stations provided 10-minute data, following a modernization of the Danish \gls{synop} station network. Although high-frequency fix-point data (sub-3-hourly) is not directly assimilated in \gls{danra}'s \gls{3dvar} system, it has proven invaluable for validation and for gap-filling when full-hour observations are unavailable. Additionally, these local datasets offer greater precision, with four-digit coordinate accuracy compared to the two-digit precision historically used in reporting observations via \gls{bufr} messages, along with corrections to station positions.  

To enhance observational input over the Danish area, data from multiple sources were evaluated. The final in-situ observation dataset was compiled from three distinct sources. First, \gls{bufr} data from the \gls{mars} archive at \gls{ecmwf} were used. Second, observations from the \gls{dmi} operational observation database (DMIdb) were compiled. Finally, archived \gls{bufr} data originally used for the operational \gls{nwp} forecasts at \gls{dmi} since mid-1994 - though initially containing some gaps - were also extracted. While these three datasets should ideally be identical, practical comparisons reveal significant differences in both coverage and content. To ensure maximum coverage with optimal quality, a rigorous quality assurance process, including manual inspections, was implemented.

 The inspection identified a substantial number of deficiencies in observations from different datasets. Consequently, significant effort was dedicated to selecting the highest-quality observational data for both assimilation and validation. For specific parameters, \gls{era5} time series proved valuable in identifying anomalies clearly unrelated to weather regime changes.

 While the different data sources often contain overlapping observations, a multitude of inconsistencies can arise. The origins of these discrepancies vary, stemming from manual errors, rounding errors, or other unidentified factors. Some of the most commonly detected errors include:
\begin{itemize}
    \item pressure offsets by 1, 10, or 100 hPa.
    \item temperature offsets by 1 or 10 K.
    \item incorrect sign of temperatures recorded in Celsius.
    \item wind speeds off by a factor of 10.
    \item wind (speed and/or direction) frozen for a longer period.
    \item wrong barometer height used in conversion from surface to \gls{mslp}.
    \item succeeding sets of observations being identical.
\end{itemize}

In the early part of the reanalysis period, prior to 2000, several cases with poor \gls{mslp} due to the absence of 2-meter temperature occurred. These data have been discarded. In the same period, the datasets include occurrences where the same data (pressure, relative humidity, temperature, and wind) appear in 2 consecutive reportings, one or three hours apart. Such sightings are extremely unlikely, and in many cases, they occur in one or two of the three datasets, but not in all of them. In such cases, data from the record exhibiting the expected behaviour was selected. It can also be noted that the quality of resolution of wind direction is much better in the last half of the reanalysis period (1\,degree uncertainty) than in the first, where uncertainties are 30 or 10\,degree. 

\begin{figure}[t]
\centering
    {{\includegraphics[width=\textwidth]{"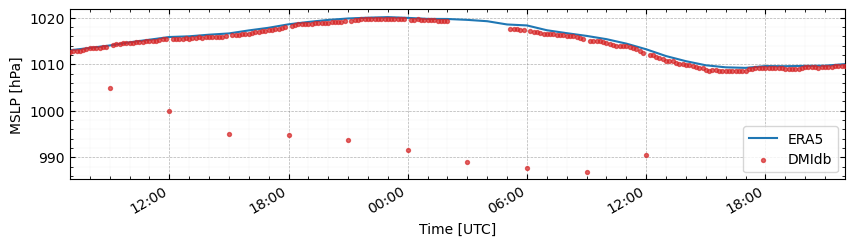"} }}
   \caption{Time series of \gls{mslp} from DMIdb (red) and ERA5 (blue) for station 06118 (S\o nderborg airport) during an episode between 11 and 12 March 2009.}
    \label{fig:pres_6118_200903}%
\end{figure}

Figure \ref{fig:pres_6118_200903} presents an example of examining an observation record of \gls{mslp} recorded at 10-minute intervals from DMIdb for Sønderborg airport station (06118) in south Denmark during March 11-12 2009, where several data points need to be removed. During the inspection step, the corresponding time series from \gls{era5} is used as background reference. As shown in the figure, while the measurement data for the selected time slice in DMIdb generally correspond well to the \gls{era5} time series, there is a segment of the DMIdb \gls{mslp} time series that exhibits significant drops in magnitude and at hourly intervals. Notably, the anomaly occurs at minute 00 in these cases, whereas the values at other minutes are more aligned with each other and \gls{era5}. Further, during hours 02 and 05 Z of March 12, 2009, only a single, erroneous observation is available. For a record as the one presented in Figure \ref{fig:pres_6118_200903}, the data assimilation routine would first remove all observations except those at analysis time (full hour). In this case, the remaining observation would be erroneous. The algorithm would most likely discard it due to a large discrepancy between the initial guess and the observation in \gls{nwp}. However, by manually removing the obviously erroneous observations, the assimilation algorithm, using a three-hour assimilation window, will include temporally close observation points for analysis, thereby fully exploiting the high-frequency temporal observations.

Thus, while the data assimilation algorithm within the \gls{nwp} model would flag erroneous observations in such records and not use them for the assimilation, information is lost in such cases. The quality assurance of \gls{danra} ensures that this information is retained and used by the reanalysis system.

\begin{figure}[t]
\centering
    {{\includegraphics[width=\textwidth]{"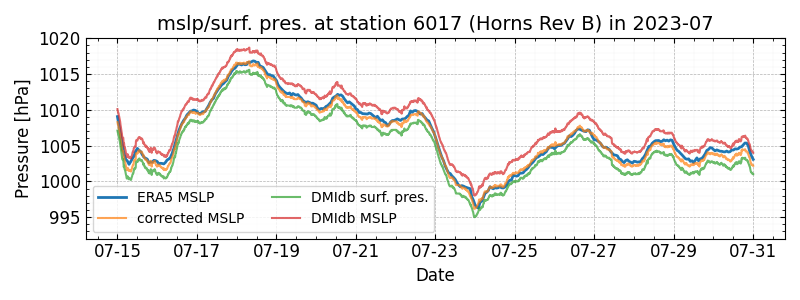"} }}
   \caption{Time series of DMIdb surface pressure (green) and DMIdb MSLP (red), corrected MSLP (orange) and ERA5 MSLP (blue) for 06017 (Horns Rev B) during second half of July 2018.}
    \label{fig:pres_6017_201807}%
\end{figure}

The persistent use of inaccurate, sometimes erroneous, barometer heights in the derivation of \gls{mslp} from surface pressure measurements is a commonly encountered quality issue in data records generated for the Danish weather observation network. During data inspection, potential quality problems associated with incorrect barometer heights can be detected through comparison with model equivalents such as \gls{era5}. Typically, wrong barometer heights get exposed in such comparisons through a systematically large bias accompanied by a relatively small standard deviation error. To correct for such error, barometer heights may be estimated by minimizing long-term average differences in \gls{mslp} to that of the \gls{era5} time series. The inferred barometer height can then be used to derive a corrected \gls{mslp} from surface pressure. One such example of data rescue is shown in Figure \ref{fig:pres_6017_201807}. The green curve and red curves represent the observed surface pressure corresponding to the original \gls{mslp} from DMIdb. The orange curve is re-derived \gls{mslp} data using the corrected barometer height, and the blue is the \gls{era5} reanalysis nearest neighbour \gls{mslp}.

\subsection{Creation of DANRA Zarr archive}\label{sec:zarr}
The complete \gls{danra} dataset was created as GRIB files (GRIB version 1), using the tools available as part of the integrated HARMONIE-AROME forecasting system. The conversion efforts by e.g. Google and Amazon have shown that to provide a dataset that is widely accessible without technical domain knowledge barriers and ready to use \citep{carver2023arco}, datasets need to be converted and aligned with modern infrastructure. Inspired by this successful approach, \gls{danra} is released officially as analysis-ready, cloud optimized reanalysis product making conversion efforts by third-parties unnecessary. As a consequence the GRIB files have been converted to CF-compliant Zarr \citep{davis_2024}. The Zarr format enables more flexible access, facilitates easier data analysis, and provides efficient parallel and cloud computing capabilities \citep{Gowan2022zarr}. Additionally, it is more suitable for machine learning workflows \citep{Nakamura2025}. As an example, 30 years of wind speed (10 metre $u$ and $v$ components) can be loaded from a network drive, and combined to compute the wind speed distributions for a single grid point in 38 $\pm$ 10 seconds (average $\pm$ standard deviation of 7 runs). For comparison, loading just a \textit{single} wind component for just \textit{one} month from GRIB files on the same network drive takes 419 $\pm$ 14 seconds (average $\pm$ standard deviation of 7 runs). Loading the GRIB files, however, also requires decompressing tar files and writing index files for the GRIB files. It would be faster if index files were pre-computed and GRIB files were uncompressed.

The conversion was done using gribscan \citep{Kolling_gribscan} to create index files for the GRIB messages, and load them into Xarray \citep{Hoyer_xarray_N-D_labeled_2017}. From there, they were saved as Zarr datasets \citep{zarr_python} with chunking in both space and time ($(t,x,y)=(256,295,263)$, average size $\approx 50$Mb). The optimal chunking depends heavily on the use case. For applications requiring a single snapshot of the entire domain, users should not have to fetch and read too many time steps. Conversely, if a time series for a single point is needed, users should not have to fetch and read too many grid points. By chunking in both space and time, we try to strike a balance between these two considerations.

For more information on the structure of Zarr files, data content, usage examples, and access information, please refer to the links provided in the data availability statement (section \ref{sec:dataavailability})

\section{Products evaluations}

Kilometre-scale reanalysis offers distinct advantages over coarse-resolution products, including a more accurate representation of local weather features, an improved ability to resolve extreme events, and a higher spatial resolution for regional climate applications. During the production of \gls{danra}, rigorous quality assurance measures were implemented, including regular monitoring and inter-comparison with ERA5—both in terms of fit to in-situ observations and through case studies about high-impact weather events (Yang et al, 2021). In this paper, we present verification statistics to demonstrate DANRA’s superior alignment with surface observations and assess its performance during some of the most extreme weather events.

\subsection{Verification against station observations}
In the following, we evaluate the agreement between \gls{danra} reanalysis and surface observations of \gls{t2m} and \gls{s10m} across Denmark, using \gls{era5} as a benchmark. Following the usual practice in operational NWP verification at \gls{dmi}, this verification utilizes only data from well-established Danish SYNOP stations, which have a long history of reporting to the \gls{gts}. The analysis covers the entire 34-year reanalysis period. During the earlier years, the Danish SYNOP network was relatively sparse, with infrequent daily reporting.  The data from this period also exhibits various deficiencies, including incorrect units, frozen values, and metadata errors. Since the late 1990s, however, station density and reporting frequency have improved significantly, leading to better data quality and quantity.

\begin{figure}[t]
\centering
    \includegraphics[width=14.5cm]{"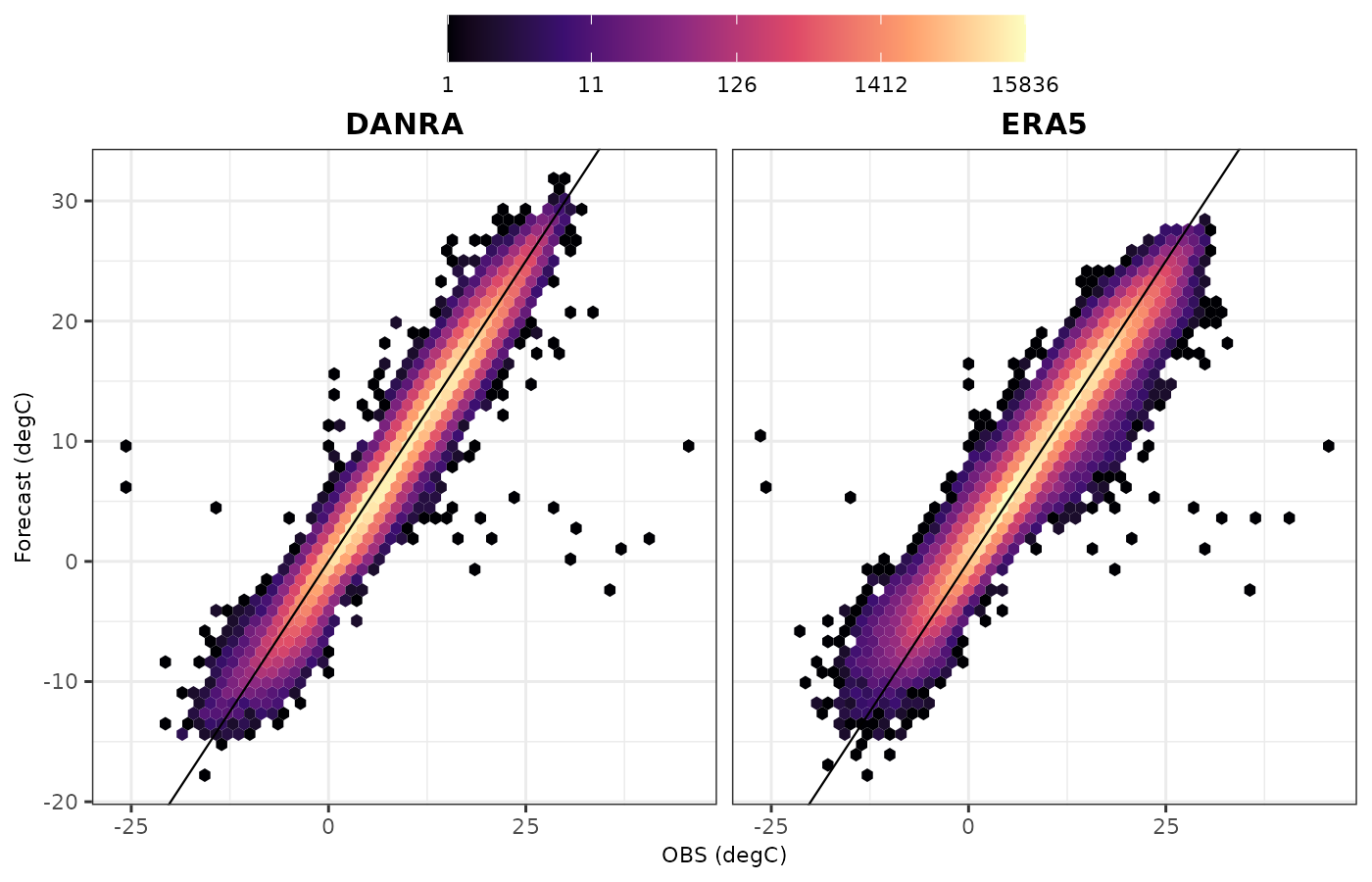"}
    \caption{Scatter plot of 3-hourly analyses for 2m temperature (\gls{t2m}) by \gls{danra} (left) and \gls{era5} (right) versus SYNOP observations over Denmark during 1990–2023. The color scale shows the density of points.}
    \label{fig:scatterT2m}
\end{figure}

\begin{figure}[t]
\centering
    \includegraphics[width=14.5cm]{"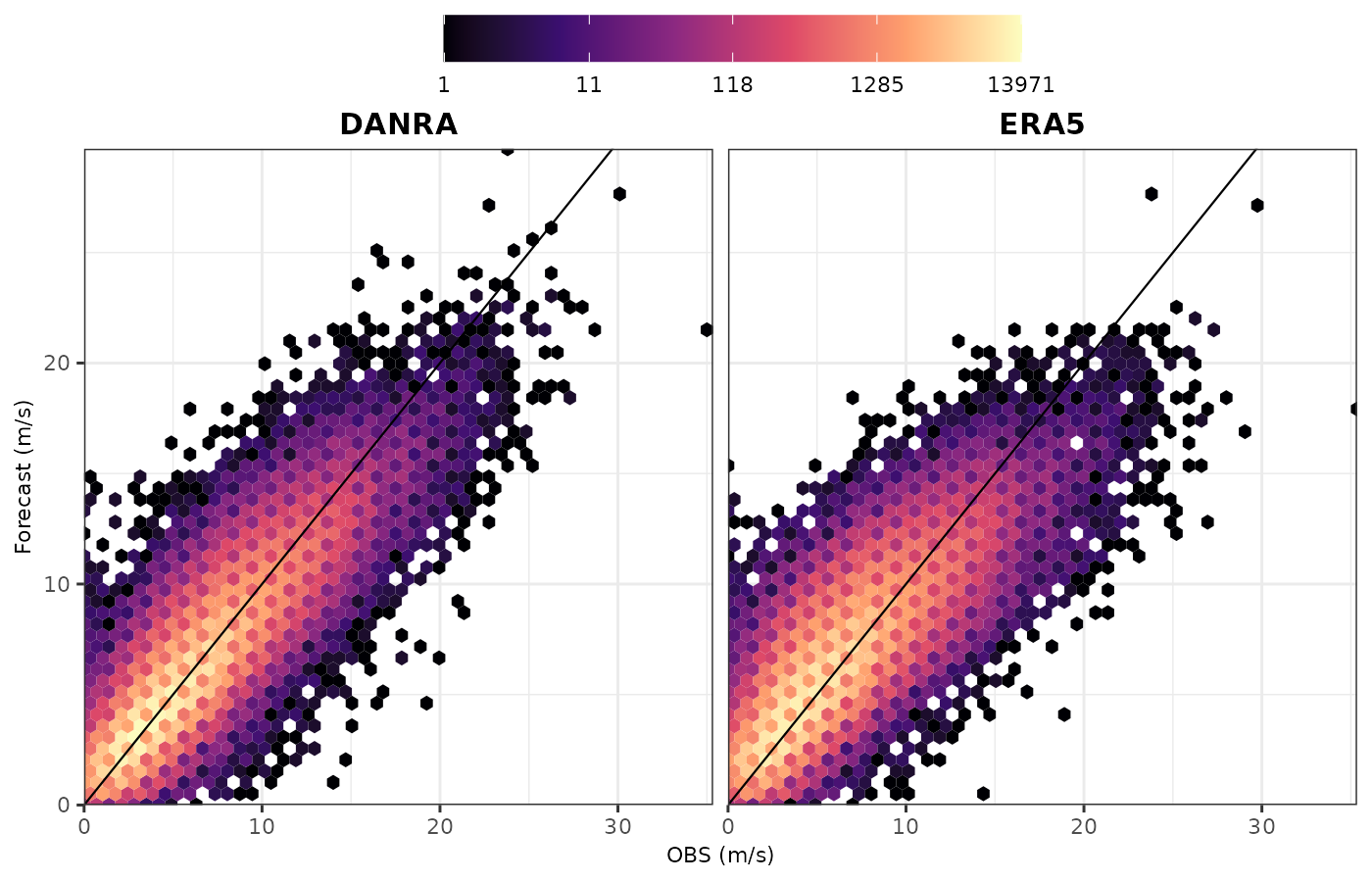"}
    \caption{Scatterplot of 3-hourly 10m wind speed (\gls{s10m}) analyses by \gls{danra} (left) and \gls{era5} (right) versus SYNOP observations over Denmark during 1990–2023. The color scale shows the density of points.}
    \label{fig:scatterS10m}
\end{figure}

To assess overall performance, we begin by comparing modelled and observed near-surface temperature and wind speed. Figure~\ref{fig:scatterT2m} and ~\ref{fig:scatterS10m} show scatter plots of \gls{t2m} and \gls{s10m}, respectively.
In both figures, \gls{danra} exhibits a tighter distribution along the identity line compared to \gls{era5}, indicating reduced standard deviation and improved alignment with observations.
Notably, for temperature, \gls{danra} demonstrates a narrower spread, whereas \gls{era5} reveals a systematic warm bias. For wind speed, \gls{danra} effectively mitigates the persistent underestimation observed in \gls{era5}, particularly at higher wind speeds.

\begin{figure}[h!]
    \centering
    \includegraphics[width=14.5cm]{"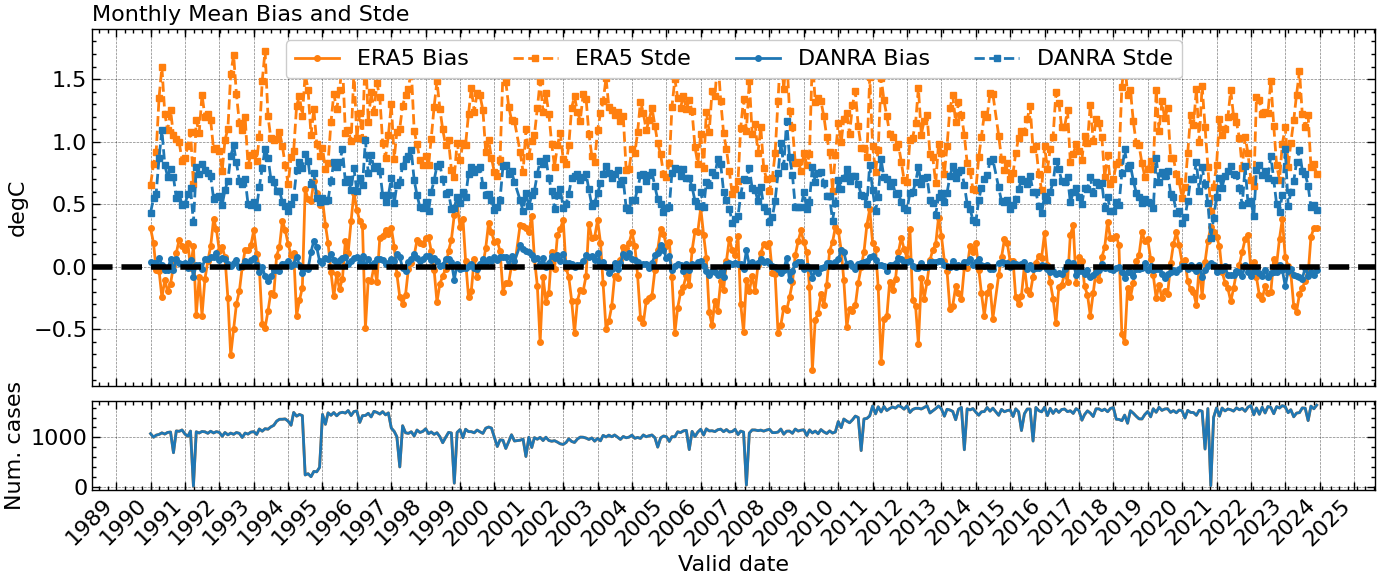"}
    \caption{Time series of monthly bias and monthly standard deviation errors for \gls{t2m} in comparison to in-situ observation for the period 1990-2023. The \gls{danra} reanalysis is shown in blue, while \gls{era5} is shown in orange. The lower panel depicts the corresponding number of matching cases for each month.}
    \label{fig:veri10}%
\end{figure}

The long-term verification statistics complement the scatter plots. Figure~\ref{fig:veri10} shows the monthly averaged time series of bias and standard deviation errors comparing fit to \gls{t2m} observations by the  \gls{danra} and \gls{era5} data sets across the entire reanalysis period (1990–2023). 

\begin{figure}[t]
\centering
    \includegraphics[width=14.5cm]{"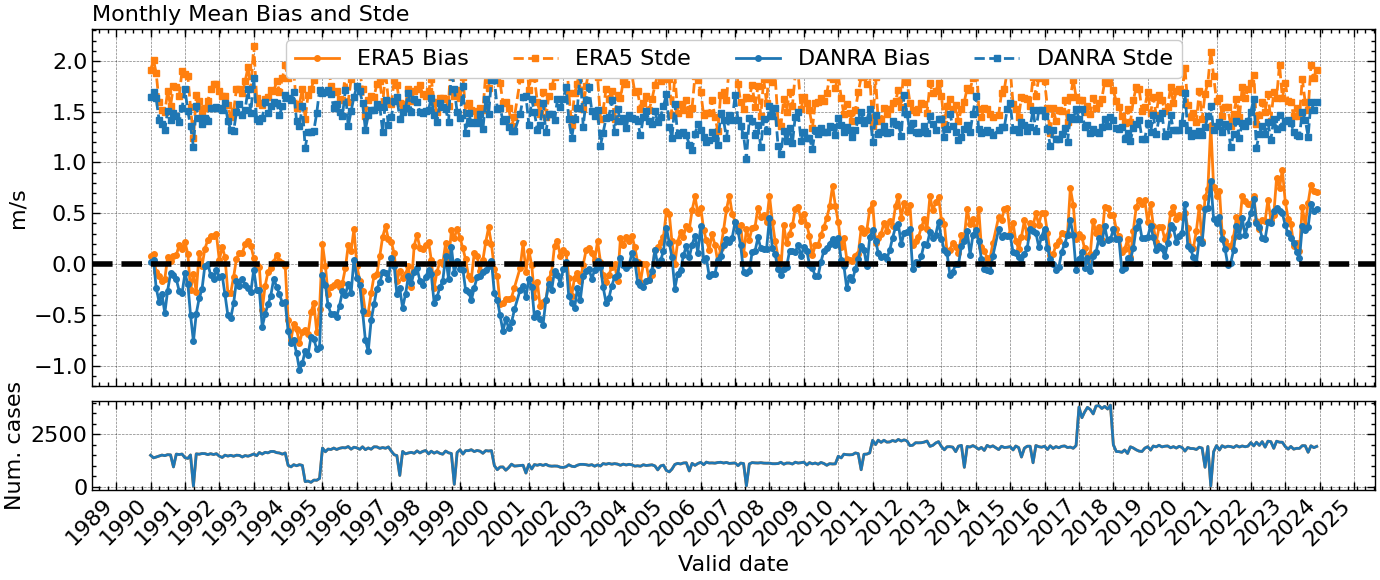"}
   \caption{Time series of monthly bias and monthly standard deviation errors for \gls{s10m} in comparison to in-situ observation for the period 1990-2023. The \gls{danra} reanalysis is shown in blue, while \gls{era5} is shown in orange. The lower panel depicts the corresponding number of matching cases for each month.}
    \label{fig:veri11}%
\end{figure}

Due to its significantly higher model resolution and the assimilation of additional surface observations, \gls{danra} demonstrates a clear added value compared to \gls{era5} throughout the reanalysis period. As illustrated in the lower panel of Figure~\ref{fig:veri10}, the number of station data available for statistical analysis varies over time. This reflects the evolving nature of observation networks and data availability, a common challenge in climate reanalysis, where changes can sometimes be substantial.
In the corresponding time series for \gls{s10m} as shown in Figure~\ref{fig:veri11}, \gls{danra} reveals a significant advantage in terms of standard deviation error, clearly indicating the added value of higher-resolution reanalysis compared to \gls{era5}. Regarding bias trends, both \gls{danra} and \gls{era5} exhibit a shift from a negative trend before 2005 to a positive trend in the past two decades. Occasionally, the bias error time series in Figure~\ref{fig:veri11} shows large jumps, which often coincide with significant drops in the number of sampled data available for statistical analysis.
Given the substantial variability and evolution in the number of available observations over the reanalysis period, caution is advised when interpreting trends in bias error, as depicted in both Figure~\ref{fig:veri10} and Figure~\ref{fig:veri11}.

It is important to note that, when interpreting the relative performance of reanalysis fits to station observations, surface wind observations, unlike \gls{t2m}, are not directly assimilated into the reanalysis process. Consequently, the agreement between reanalysis and independent wind measurements is more strongly influenced by representation errors. As a result, the correspondence between modelled and observed wind speed time series is less precise than that for \gls{t2m}.

\subsection{\gls{danra}'s representation of extreme weather events}

\begin{figure}[t]
\centering
    {{\includegraphics[width=\textwidth]{"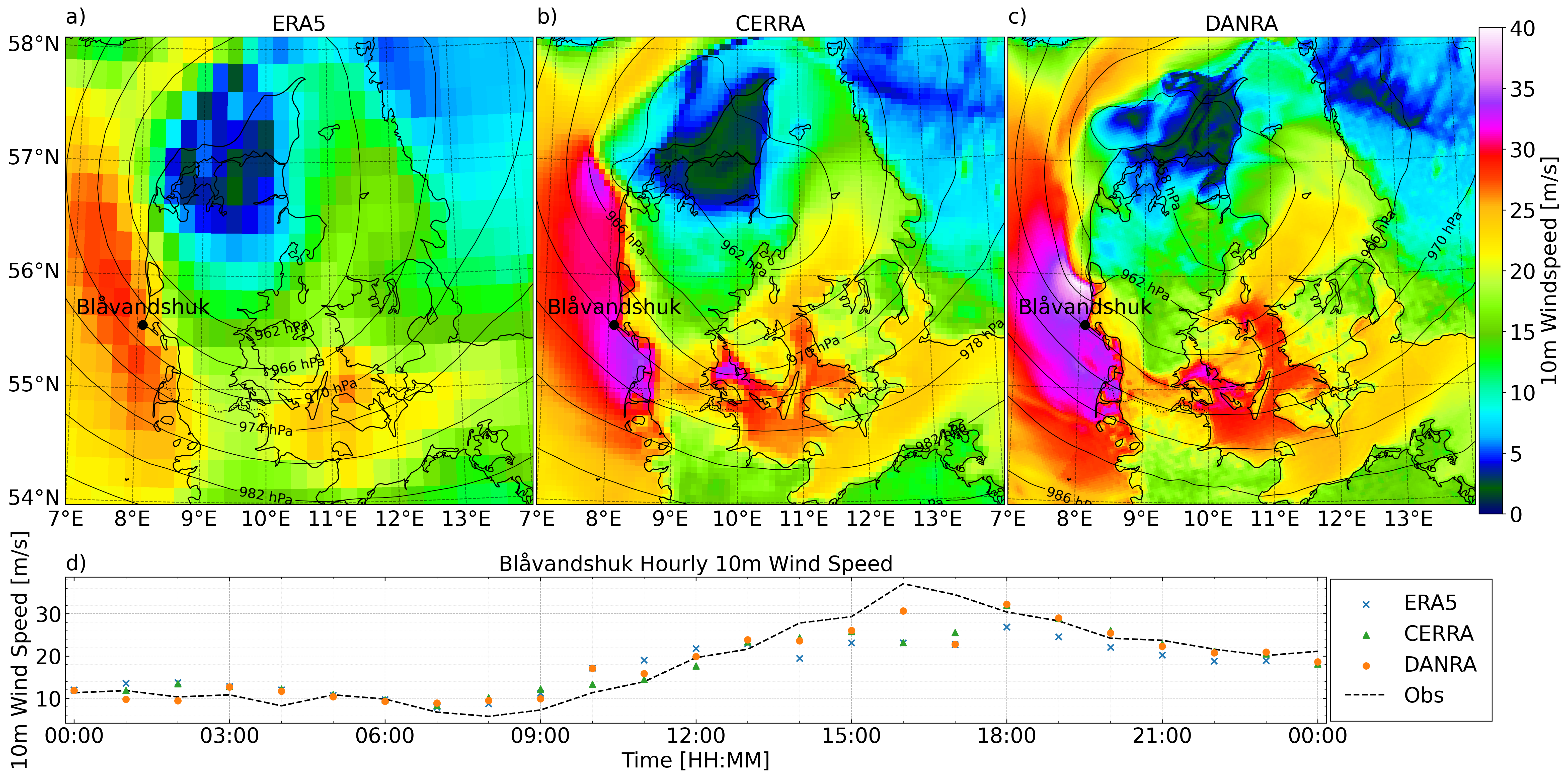"} }}
    \caption{\textbf{10 m Wind Speed and Mean Sea Level Pressure on December 3, 1999, 18:00 UTC.}  Comparison of \gls{era5}, \gls{cerra}, and \gls{danra} reanalyses during the December 1999 hurricane, one of the most severe extratropical cyclones to affect Denmark in modern times. The top panels show spatial distributions of 10 m wind speeds with mean sea-level pressure contours at 4 hPa intervals. The bottom panel shows hourly 10 m wind speeds measured at the Blåvandshuk \gls{synop} station (black dashed line) together with the corresponding values from \gls{era5} (blue), \gls{cerra} (green), and \gls{danra} (orange).}
    \label{fig:wind_case}
\end{figure}

For regional climate reanalysis, \gls{danra} must demonstrate a strong capability to accurately represent major weather events and extreme climate conditions throughout the reanalysis period. In \citet{yang2021}, the \gls{danra} reanalysis was evaluated for its representation of historical storms and heavy precipitation events. In this study, we present \gls{danra}’s performance in simulating three extreme events: a storm, a historical heatwave, and a flash flood. To highlight the added value of km-scale reanalysis, we also compare \gls{danra} with both the 5.5 km \gls{cerra} reanalysis data and the 31 km global \gls{era5} reanalysis data, both of which are available through the \gls{cds}. Note that in these case studies, while the \gls{t2m} and \gls{s10m} data at (3-hourly) analysis time  (00:00 UTC, 03:00 UTC...) are extracted from reanalysis, for hourly data at non-analysis time (01:00 UTC, 02:00 UTC...), short-range forecasts of 1- to 2-h have been extracted. For precipitation, accumulated precipitation from the twice-daily reanalysis forecast steps has been extracted.

\subsubsection{Historical storm on December 3rd, 1999}

One of the most powerful cyclones in Denmark’s modern history, the December 3, 1999 storm, serves as our case study for extreme wind events. This intense extratropical cyclone, marked by hurricane-force sustained winds, underwent rapid deepening over the North Sea before tracking eastward and crossing Denmark from the late afternoon of December 3 into the early hours of December 4. At 18:00 UTC, the storm reached its peak intensity over Denmark, with a minimum surface pressure of 952 hPa near Skive in North Jutland \citep{voldborg2000}. During its passage, hurricane-force average wind speeds were recorded at multiple locations. Raw data reveal that both Rømø (Wadden Sea) and Røsnæs (western Zealand) registered an extreme average wind speed of 41.2 m/s. At Rømø, wind gusts reached 51.4 m/s before the instrument was destroyed. Even inland, the storm produced exceptional winds—most notably in Skrydstrup (Southern Jutland), where a mean wind speed of 31.4 m/s approached hurricane force, despite the station’s inland location.

In addition to the extreme winds, the event also generated a major storm surge, with water levels rising to 5.0 meters above normal—nearly breaching the dikes in the Wadden Sea. Fortunately, the peak wind speeds coincided with an astronomical low tide, which significantly reduced the potential impact.

Figure \ref{fig:wind_case}a shows the \gls{era5} analyses for \gls{mslp} and 10m wind speed at 18 UTC on December 3, 1999 (upper panel). Notably, no 958 hPa isobars are present, and hurricane-force winds (>32.6 m/s) are absent—even over coastal waters. The maximum mean wind speed in \gls{era5} reaches only 28.5 m/s, corresponding to a "violent storm" scale on the Beaufort scale. This simulated peak is approximately 30 \%  lower than the observed peak winds at coastal stations along western Jutland. Similarly, the 5.5 km \gls{cerra} reanalysis (Fig \ref{fig:wind_case}b), also appears to underestimate the storm's intensity, with a simulated surface pressure that is too shallow. In contrast, the \gls{danra} reanalysis (Fig. \ref{fig:wind_case}a), with a minimum pressure of 954 hPa, provides a much closer match to observation in both magnitude and spatial positioning. 

In terms of wind speed, \gls{danra} data also demonstrates significantly improved alignment with observations. At 18 UTC, both \gls{cerra} and \gls{danra} feature strong pressure gradients, resulting in a markedly higher wind speed than \gls{cerra}. Specifically, in \gls{danra}, coastal regions just north of Blåvandshuk include several grid cells (highlighted in whitish colors) with wind speeds approaching 40 m/s, peaking at 40.2 m/s—though this area lacks weather stations for direct validation. Where stations are available, mean wind speeds consistently range from 33 to 38 m/s, closely matching in-situ measurements from DMI’s automated weather stations. 

In Figure \ref{fig:wind_case} d), hourly measurement time series from Blåvandshuk are compared with the nearest grid cells data from \gls{era5}, \gls{cerra}, and \gls{danra}. All three reanalyses track the storms evolution well until the peak period between 15:00 and 18:00 UTC, when their performance diverges. While observations show a clear wind speed peak at 16:00 UTC, \gls{era5} and \gls{cerra} time series depict a slower increase, reaching their maximum at 18:00 UTC. In contrast, the \gls{danra} time series exhibits a double-peak pattern, with wind speed exceeding 30 m/s at both 16:00 UTC and at 18:00 UTC. The dip between these peaks in the \gls{danra} series likely corresponds to a shift in wind direction from south-westerly to north-westerly, in accordance with the process as described in \citep{voldborg2000}.

The absence of this double-peak pattern in the observations suggests that the feature captured by \gls{danra} may be small-scale and challenging to simulate accurately. Nevertheless, \gls{danra} provides the best overall match to the observed time series.

\subsubsection{Historical heat wave, July 20th 2022}

\begin{figure}[t]
\centering
    {{\includegraphics[width=\textwidth]{"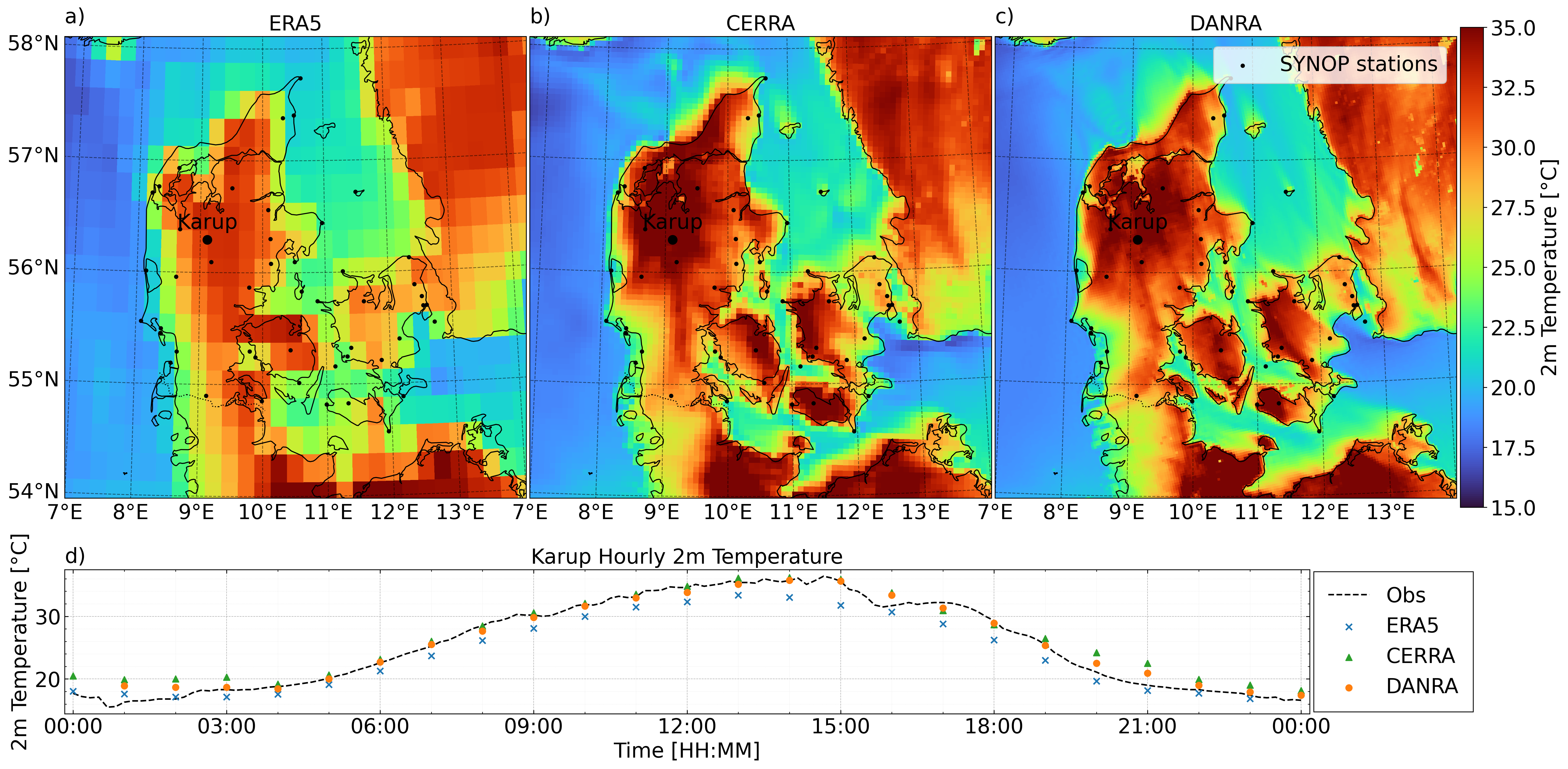"} }}
\caption{\textbf{2 m Temperature • July 20 2022, 15:00 UTC.}  
Comparison of \gls{era5}, \gls{cerra}, and \gls{danra} reanalyses during the historic July 20 2022 heatwave, which set the highest measured Danish temperature record of 36.7~$^\circ$C at Karup.  The top panels show spatial distributions of 2 m temperature. The bottom panel shows hourly 2 m temperatures measured at the Karup SYNOP station (black dashed line) together with the corresponding values from \gls{era5} (blue), \gls{cerra} (green), and \gls{danra} (orange). The small black dot in the figures indicate positions of the official Danish SYNOP observation stations.}
    \label{fig:temp_case}%
\end{figure}

The ability to reproduce temperatures correctly is one of the most crucial elements in \gls{nwp}. We examine here a historically hot day in July 2022 and compare \gls{danra} data with \gls{cerra} as well as \gls{era5}.

On July 20, 2022,  an intense surge of hot air from the south and southeast swept across Northern Europe, pushing temperatures in Denmark to unprecedented levels. At over 10 official weather stations operated by  \gls{dmi}, temperatures soared above 35~$^\circ$C. The highest recorded temperature, 36.7°C, was observed at Karup station in central Jutland, marking the highest temperature ever measured in Denmark.  The heatwave was widespread, with temperatures exceeding 35~$^\circ$C  across Jutland, Funen, and Lolland. Due to the southeasterly winds, the most extreme temperatures occurred in the northwestern regions of each area, where the air had traveled over extensive sun-warmed land, maximizing the transfer of sensible heat from the surface.
In contrast, coastal stations with onshore winds recorded the lowest temperatures, as the surrounding waters averaged just under 20°C. Downwind of fjords, bays, and other bodies of water, cooler air extended inland, while offshore winds carried warm air out to sea before it could cool. Even areas near lakes exhibited slightly lower temperatures. Accurately reproducing these spatial temperature variations in numerical weather prediction (NWP) models requires sufficient horizontal resolution.

Figure \ref{fig:temp_case} shows the 2 metre temperature reanalysis by \gls{era5} (a), \gls{cerra} (b), and \gls{danra} (c) at 15:00 UTC on 20 July 2022, the time point when maximum temperatures have been simulated. In figure \ref{fig:temp_case}d, time series of the 2 metre temperature from observation in station Karup, mid-Jutland in Denmark, with a 10 min interval, have been plotted,  together with hourly model data from the nearest grid boxes from \gls{era5}, \gls{cerra}, and \gls{danra}. For this historical heat wave case, simulations by both regional reanalyses, \gls{cerra} and \gls{danra}, show an excellent correspondence in comparison to the measurement record. As shown in Fig. \ref{fig:temp_case}, for that day, a relatively widespread region in Denmark has had temperatures over 35 degrees, according to simulations by both \gls{cerra} and \gls{danra}. Temperature of over 37~$^\circ$C is simulated over a narrow, elongated band in western and northwestern Jutland with peak value in \gls{danra} reaching as high as 37.8~$^\circ$C. Although this cannot be directly verified due to a lack of 2-metre temperature measurements in that area, it testifies to a historically hot day with exceptionally high temperatures. On the same day, 40.1~$^\circ$C was recorded in Hamburg in northern Germany.

\gls{era5} also simulates a very warm day with widespread temperatures over 32~$^\circ$C. The highest simulated grid value is also found in the northern part of central Jutland, where, according to \gls{era5}, the temperature is 34.7~$^\circ$C. This value is lower than what has been measured in many places at official weather stations, and 3.1~$^\circ$C lower than the highest grid value that appears in \gls{danra}. For station Karup, as shown in Figure \ref{fig:temp_case}d, the \gls{era5} data appears to suffer a cold bias of around 5 degrees close to the peak of the event.

It is evident that, owing to the coarser horizontal resolution, the same level of detail shown in kilometer scale \gls{danra} is not visible with \gls{era5}. The temperature footprint of the island of Bornholm, in the middle of the Baltic Sea, is not observable in \gls{era5}. In general, the temperature gradient between sea and land in \gls{era5} is irregular and not as realistic as in \gls{danra}.

While comparing model simulations such as those in Figure \ref{fig:temp_case} against observations on fine-scale variability of temperature fields as revealed by the kilometer-scale model \gls{danra}, it may be interesting to reflect on the limitations of the existing observation network in representation of important weather where small-scale features may be dominating. In figure \ref{fig:temp_case} (a)-(c), station positions of the Danish \gls{synop} observation network have been marked by small black dots to illustrate the distribution of the surface observation network. While these variability appear to contain many small scale details, presumably as consequence of forcing of various origins including both external and local conditions, it is evident that the existing surface observation network as shown in the figure \ref{fig:temp_case}, owing to the limitation in station density, is far from sufficient to resolve/represent all the essential scales of importance affecting weather and climate. 

\subsubsection{Extreme cloudburst over south Jutland, August 20th, 2007}

\begin{figure}[t]
\centering
    {{\includegraphics[width=\textwidth]{"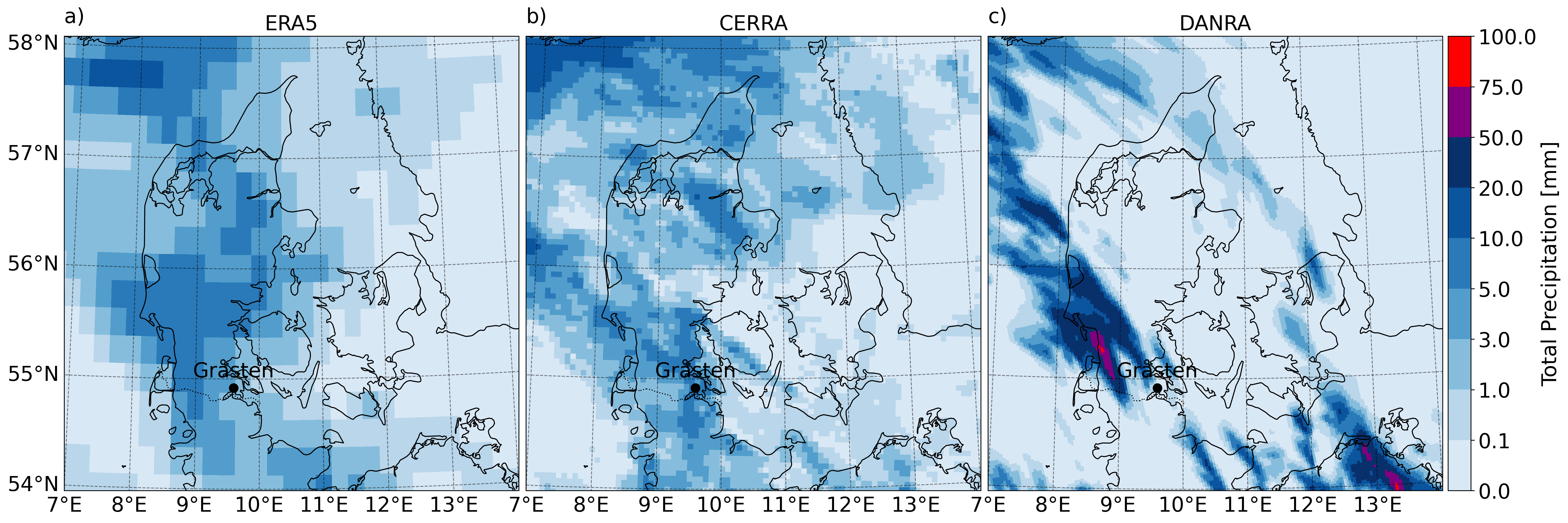"} }}
\caption{\textbf{Accumulated Precipitation • August 20 2007, 15:00–21:00 UTC.}  
Comparison of \gls{era5}, \gls{cerra}, and \gls{danra} reanalyses during the 20 August 2007 cloudburst, an extreme precipitation event recorded in Denmark, with more than 140 mm falling in 1.5 hours over Gråsten.  
Each panel shows the spatial distribution of precipitation totals accumulated over the 6-hour period ending at 21:00 UTC.}

    \label{fig:precip_case}%
\end{figure}

On the evening of August 20, 2007, an extremely violent cloudburst struck the south-eastern part of Jutland, where enormous amounts of rain fell locally within a short time. At Gråsten, a now decommissioned \gls{dmi} station recorded 142 mm of rain in one and a half hours, and a subsequent analysis of radar-estimated precipitation concluded that the rainfall intensities were exceptionally high. During this episode, the maximum precipitation within 10 minutes was 53 mm, and 111.3 mm within 30 minutes. For comparison, according to \gls{dmi} definition, a cloudburst in Denmark is defined as more than 15 mm in 30 minutes. Thus, at the peak of this event, the precipitation intensity was more than ten times the cloudburst threshold. The consequences of the extreme rainfall included a road being washed away by the tremendous water masses and a railway line becoming undermined, so that the tracks were left hanging freely several meters above the ground. Fortunately, no one was injured during the extreme event.

Figure \ref{fig:precip_case} (a), (b), and (c) present the precipitation output from \gls{era5}, \gls{cerra}, and \gls{danra}, respectively, during the period between 15 and 21 UTC on the evening of the cloudburst event. It is shown that \gls{danra} produces a clear convective precipitation signal, with maximum simulated total precipitation exceeding 75 mm over an elongated area about 30 km northwest of Gråsten, the approximate center of the event. Despite a slight phase shift in the central location of the simulated convection, given the inherent challenges in forecasting convective storms—both temporally and spatially, it appears to be a great success that the kilometer-scale \gls{danra} succeeds in reproducing the signature of intense convective activity. In contrast, the heavy convection situation appear absent in both of the other two reanalyses \gls{era5} and \gls{cerra}. The simulated precipitation amounts with these coarser-resolution reanalyses are generally very low, with accumulated rain well below double digits in millimeters. This highlights a key constraint in simulating events driven by \gls{dmc}, where kilometer-scale spatial resolution is an essential requirement to ensure that reanalysis projects accurately represent localized and convective precipitation.

\glsresetall
\section{Summary and outlook}

Over recent years, atmospheric reanalysis has become an indispensable tool in climate research and monitoring. By applying data assimilation tools within state-of-the-art operational \gls{nwp} systems, reanalysis generates long-term, gridded datasets of past atmospheric states. Unlike traditional climate statistics derived from observational records, reanalysis datasets offer the distinct advantages of spatial completeness, temporal continuity, and consistency—attributes that are particularly valuable for climate studies due to the use of a fixed model version. This study introduces \gls{danra}, a 34-year climate reanalysis dataset focused on the Danish region, generated by a high-resolution (2.5 km grid-spacing) regional atmospheric reanalysis system. It provides gridded, gap-free climate information with unprecedented detail and fidelity for use in climate research and societal applications in Denmark.

The \gls{danra} reanalysis system integrates components from both the operational \gls{nwp} model at \gls{dmi} and the \gls{carra}, both of which are based on HARMONIE-AROME 40h1. During the development phase, the system and its production infrastructure, including modules for monitoring, quality assurance, and post—processing, were established. The \gls{danra} reanalysis was carried out on the \gls{ecmwf} \gls{hpcf}, leveraging existing infrastructure from regional reanalysis projects such as \gls{carra} and \gls{cerra}, particularly for observation data management. 

Extensive work on local observation data acquisition, quality assurance, and data rescue has been conducted to enhance observation input in the reanalysis. Quality assurance for \gls{danra} involves validation against in-situ observations, as well as inter-comparison to the corresponding datasets from other reanalysis projects, with the underlying assumption that high-resolution regional reanalyses should outperform coarser resolution products in representing key weather parameters (e.g., near-surface temperature, wind, and humidity) due to their enhanced ability to resolve local surface effects. As described in this paper, both the validation statistics and case studies confirm this expectation, demonstrating the added value of \gls{danra}’s high resolution in representing the weather and climate characteristics important to various applications, such as climate research and adaptation.

A fundamental requirement for climate reanalysis datasets is the precise depiction of weather extremes, which are expected to become more frequent and intense due to global warming. In Denmark, high-impact events such as flash floods, storms, and heatwaves are heavily influenced by local surface conditions, making their accurate simulation highly dependent on model resolution. 

Analyses of severe storm events over the past three decades—many of which reached hurricane-scale intensity—demonstrate that \gls{danra} (Danish Reanalysis) more faithfully reproduces strong wind conditions compared to coarser-resolution reanalyses. As a cloud-resolving model, \gls{danra} also excels in simulating heavy precipitation events with strong local characteristics, whereas coarser models like \gls{era5} often underestimate or entirely miss these events.

Inter-comparisons among \gls{era5}, \gls{cerra}, and \gls{danra} in case studies of high-impact weather extremes—including wind, temperature, and precipitation—reveal a clear pattern: higher grid resolution consistently improves the accuracy of event representation. These findings underscore the effectiveness of the kilometer-scale HARMONIE-AROME model in capturing local-scale weather features, such as severe convective events, and affirm \gls{danra}’s role as a valuable tool for regional climate research in Denmark.

With its high spatial grid-spacing of 2.5 km and temporal coverage spanning more than three decades, the \gls{danra} dataset provides a unique foundation for advancing research on data-driven models. In particular, it supports the development and evaluation of \gls{lam}-based machine learning frameworks that rely on the accurate representation of fine-scale atmospheric processes, such as coastal circulations, convective initiation, and orographic precipitation. These detailed structures, often unresolved or poorly represented in global reanalyses, are crucial for enhancing predictive skill at regional scales and for exploring new methodologies in modeling.

%% The following commands are for the statements about the availability of data sets and/or software code corresponding to the manuscript.
%% It is strongly recommended to make use of these sections in case data sets and/or software code have been part of your research the article is based on.

%\codeavailability{TEXT} %% use this section when having only software code available

%\dataavailability{TEXT} %% use this section when having only data sets available

%\codedataavailability{TEXT} %% use this section when having data sets and software code available

\section{\gls{danra} data availability}\label{sec:dataavailability}

\gls{danra} reanalyses generate gridded data for a large number of weather and climate variables, both at 3-hourly analysis time. Some of the parameters are defined on height or pressure levels, the rest are on a single level.

The \gls{danra} reanalysis data, available at three-hour intervals at 00:00, 03:00, 06:00, ..., and 21:00 UTC daily (analysis data), can be accessed from an Amazon S3 object store. The data is split into three different Zarr datasets:
\begin{itemize}
\item \href{s3://dmi-danra-05/height_levels.zarr}{s3://dmi-danra-05/height\_levels.zarr}
\item \href{s3://dmi-danra-05/single_levels.zarr}{s3://dmi-danra-05/single\_levels.zarr}
\item \href{s3://dmi-danra-05/pressure_levels.zarr}{s3://dmi-danra-05/pressure\_levels.zarr}
\end{itemize}

Fetching one of these datasets with Python and xarray using simple loading is as simple as
\begin{verbatim}
ds_danra_sl = xarray.open_zarr(
    "s3://dmi-danra-05/single_levels.zarr",
    consolidated=True,
    storage_options={
        "anon": True,
    }
)
\end{verbatim}

Efforts are currently underway to prepare the \gls{danra} forecasts for public release. Once available, the dataset will provide hourly temporal resolution. Updates and further details will be posted on the \gls{danra} documentation website \href{https://dmidk.github.io/danradocs}{https://dmidk.github.io/danradocs}, which already includes information on the Zarr file structure, data access, and usage examples.

\begin{acknowledgements}
\gls{danra} project is supported by the Danish \gls{nckf}. \gls{danra} has also benefited from the system and observation data infrastructure in the \gls{era5}, \gls{carra}, and \gls{cerra}. Colleagues at the DMI Weather Research Department, especially Irene Livia Kruse, Bent Hansen Sass, Rune Carbuhn Andersen, and Julia Sommer, are acknowledged for providing various support during different stages of the \gls{danra} project. Eoin Whelan from Met Éireann is acknowledged for providing GRIB data extraction scripts from the Met Éireann ReAnalysis (MERA). THE EUMETSAT ROM-SAF team is acknowledged for making available the ROM-SAF \gls{icdr} data with reprocessed Radio Occultation bending angle for assimilation in \gls{era5}. 
\end{acknowledgements}

%% REFERENCES

\bibliographystyle{copernicus}
\bibliography{references}

\end{document}